\documentclass[aps,pre,reprint,superscriptaddress,amsmath,amssymb,showpacs,floatfix]{revtex4-1}
\usepackage{graphicx}
\usepackage{bm}
\usepackage{color}

\begin{document}

\title{Mutual information change 
in feedback processes driven by measurement}

\author{Chulan Kwon}
\email{ckwon@mju.ac.kr}
\affiliation{Department of Physics, Myongji University, Yongin, Gyeonggi-Do,
17058,  Korea}
\affiliation{Quantum Universe Center, Korea Institute for Advanced Study, Seoul 02455, Korea}

\date{today}

\begin{abstract}
We investigate thermodynamics of feedback processes driven by measurement. Regarding system and memory device as a composite system, mutual information as a measure of correlation between the two constituents contributes to the entropy of the composite system, which makes the generalized total entropy of the joint system and reservoir satisfy the second law of thermodynamics. We investigate the thermodynamics of the Szilard engine for an intermediate period before the completion of cycle. We show the second law to hold resolving the paradox of Maxwell's demon independent of the period taken into account. We also investigate a feedback process to confine a particle excessively within a trap, which is operated by repetitions of feedback in a finite time interval. We derive the stability condition for multi-step feedback and find the condition for confinement below thermal fluctuation in the absence of feedback. The results are found to depend on interval between feedback steps and intensity of feedback protocol, which are expected to be important parameters in real experiments. 

\end{abstract}
\maketitle

\section{Introduction}
There have been long-time efforts for about 150 years to resolve the paradox of Maxwell's demon~\cite{maxwell,szilard,brillouin,landauer,bennett,leff_rex}. It states that the thermodynamic second law is violated in post-measurement process controlled by the demon's feedback from measurement. Szilard opened the era of information thermodynamics by considering a tractable prototype, later called the Szilard engine, of Maxwell's demon~\cite{szilard} which has been studied extensively up to date. The paradox was claimed that the sole effect after the completion of the Szilard engine for one cycle is entropy loss in heat reservoir, (work extraction in the absence of a cooler reservoir), which seemingly violates the thermodynamic second law. Through many efforts resolving the paradox, common perspective is that the demon should be treated as a physical memory device to measure the state of a target system and information gain/use can be defined as having the same footing as entropy, which were supported by information theory developed independently in computer  science~\cite{shannon,cover}. 

The whole information process controlled by Maxwell's demon can be divided into the measurement process with information gain and the post-measurement process with information use. For a bit-information process within a memory device, as proposed by Landauer~\cite{landauer}, one can imagine a particle moving in a double-well potential inside a memory chip. Empty memory state before measurement corresponds to localization in one well. Memory state is changed by time-varying potential that is initially perturbed by a target state of system and is set to return to the original double-well. During this process, memory state goes to a local equilibrium in either well which is read as a measurement outcome. As in real situations, measurement time is so short that system state remains unchanged during measurement. In post-measurement, the obtained measurement outcome turns into a protocol via feedback which influences the subsequent dynamics of the system. 

The first candidate of information content in measurement-feedback process was the change in Shannon entropy~\cite{shannon} of  memory device. However, it cannot deal with post-measurement process because memory state remains frozen in a local equilibrium or may be erased for the purpose of another measurement without changing the system dynamics. By using the Landauer principle, it was often claimed that heat loss in reservoir can be compensated by sufficient heat dissipation produced in the course of the erasure of memory~\cite{bennett}. However, erasure is not necessarily simultaneous with post-measurement. Indeed, the two processes are independent. 

A satisfactory thermodynamic theory integrating measurement and post-measurement has recently been developed since late 2000s~~\cite{sagawa1,sagawa2,sagawa_lecture,sagawa_recent}. The so-called information thermodynamics exploits the information science~\cite{shannon,landauer,cover} and the modern nonequilibrium principle of the fluctuation theorem~\cite{evans,jarzynski,crooks,kurchan,lebowitz,seifert,esposito}. The central notion is that system and memory be regarded as a composite one with correlation leading to an entropic contribution, called mutual information. The role of mutual information has been confirmed in feedback experiments~\cite{toyabe,koski}. Beyond resolving the paradox of Maxwell's demon, there have been many studies on feedback processes in diverse perspectives such as repeated feedback~\cite{horowitz,ponmurugan}, optimal protocol change for maximum extraction of work in feedback processes~\cite{abreu, horowitz_parrando,parrando_natphys}, cold damping~\cite{ito,munakata1}, time-delayed feedback~\cite{munakata2,rosinberg,KUP}, information flow~\cite{horowitz1,horowitz2,shiraishi}, and information engine~\cite{UHKP}.   

In this study, we investigate thermodynamic process by feedback focusing on mutual information change in time. In Sec.~2, we briefly review mutual information in order to make our work self-contained. In Sec.~3, we revisit the Szilard engine and examine thermodynamics during an intermediate period before the completion of cycle. In Sec.~4, we investigate a damping process to confine a particle excessively within a trap potential, which is operated by repetitions of feedback steps in a finite-time interval until achieving a satisfactory confinement. In Sec~.5, we summarize our study and discuss problems to be further studied.   

\section{Mutual information}

We briefly review information thermodynamics which has been covered extensively in recent literatures~\cite{sagawa_lecture,sagawa_recent}. Let $x$ and $y$ be states of system and memory, respectively, and $P(x)$ and $P(y)$ be probability distribution functions (PDFs) for the two. Then the PDF of the joint system is given by $P(x,y)=P(x|y)P(y)=P(y|x)P(x)$ where $P(x|y)$ and $P(y|x)$ are conditional probabilities. Entropy of each system is given by Shannon entropy~\cite{shannon}, $S(x)=-\ln P(x)$ and $S(y)=-\ln P(y)$, setting Boltzmann constant to be unity. Then, the entropy of the composite system is written as 
\begin{equation}  
S(x,y)=-\ln P(x,y)=S(x)+S(y)-I(x:y)~,
\label{total_shannon}
\end{equation}
where $I(x:y)$ is mutual information defined as 
\begin{equation}
I(x:y)=\ln \frac{P(x,y)}{P(x)P(y)}=\ln \frac{P(y|x)}{P(y)}=\ln \frac{P(x|y)}{P(x)}.
\label{MU_def}
\end{equation}
Mutual information is a measure for correlation between the two states and has a property $\langle I(x:y)\rangle\ge 0$~\cite{cover}. 

Recently, measurement and erasure processes were experimentally realized by using a double-well potential in optical trap~\cite{berut}. Theoretically, the formal expression for the change in total entropy of the composite system and reservoir was derived by using a multi-well potential picture~\cite{sagawa_lecture}. In reality, however, those are hidden and unobservable processes. We only concentrate ourselves to observable post-measurement process, for which the generalized second law was also derived by using the fluctuation theorem~\cite{sagawa_recent}. We will briefly summarize the derivation to make concepts and terminologies self-contained in our study. 

We consider a post-measurement process starting at time $t=0$ and ending at $t=\tau$ where measurement outcome is given from a final memory state $y$ in measurement process and determines a protocol $\lambda_y(t)$ generally dependent on time for $0\le t\le \tau$ in the subsequent dynamics. $\lambda_y(t)$ may be used for a parameter change in potential or for an effective field to drive the system. Then, the probability of the system tracing a path $x(t)$ is written as 
\begin{equation}
P[x(t);y]=P(x_0,0)P(y|x_0,0)\Pi[x(t)|x_0;\lambda_y(t)], 
\label{path_prob}
\end{equation}  
where $\Pi[x(t)|x_0;y]$ is the conditional probability of path $x(t)$ for $0<t<\tau$ starting from $x_0=x(0)$. $P(y|x,0)$ is a by-product from hidden measurement process. In reality, it can be estimated from factory-given accuracy for a measurement device such as a ccd camera. Two types of measurement probabilities are used in theoretical approaches: $P(y|x,0)=1-\epsilon$ or $\epsilon$ for binary measurement with $0\le \epsilon\le 1$ and $P(y|x,0)=e^{-(y-x)^2/(2\sigma)}/\sqrt{2\pi\sigma}$ for measurement of continuous states. We introduce an adjoint dynamics with the time-reversed protocol $\lambda^R_y(t)=\lambda_y(\tau-t)$. Then, the probability of the system tracing the time-reverse path $x^R(t)=x(\tau-t)$ in the adjoint dynamics is given by
\begin{equation}
\widehat{P}[x^R(t);y]=\widehat{P}(x_\tau,0)\widehat{P}(y|x_\tau,0)\Pi[x^R(t)|x_\tau;\lambda^R_y(t)],
\end{equation} 
where $x_\tau=x(\tau)$.

Defining $e^{R}=P/\widehat{P}$, we write
\begin{equation}
R[x(t);y]=\ln\frac{P(x_0,0)P(y|x_0,0)}{\widehat{P}(x_\tau,0)\widehat{P}(y|x_\tau,0)}+\ln \frac{\Pi[x(t)|x_0;\lambda_y(t)]}{\Pi[x^R(t)|x_\tau;\lambda^R_y(t)]}.
\end{equation}
It is well known that the second term gives heat production in heat reservoir at temperature $T$
\begin{equation}
Q[x(t);\lambda(t;y)]=T\ln \frac{\Pi[x(t)|x_0;\lambda_y(t)]}{\Pi[x^R(t)|x_\tau;\lambda^R_y(t)]},
\end{equation}
which was proven both for the Brownian motion using the Onsager-Machlup theory~\cite{onsager,kurchan,KYKP} and for jumping process in discrete states described by the master equation ~\cite{schnakenberg}. One can easily show that $R$ satisfies the integral fluctuation theorem as  
\begin{eqnarray}
\langle e^{-R}\rangle&=&\int dy\int D[x(t)] P[x(t);\lambda_y(t)]e^{-R[x(t);\lambda_y(t)]}\nonumber\\
&=&\int dy\int D[x(t)]\widehat{P}[x^R(t);\lambda_y^R(t)]=1
\end{eqnarray}
where $\int D[x(t)](\cdots)$ denotes the path integral over all paths. A direct consequence of the theorem is the inequality $\langle R\rangle\ge 0$, obtained by Schwarz inequality.

There are infinitely many kinds of $R$ depending on the choice of initial PDF of the adjoint dynamics. If we choose: $\widehat{P}(x_\tau,0)=P(x_\tau,\tau)$ and $\widehat{P}(y|x_\tau,0)=P(y|x_\tau,\tau)$, $R$ becomes the change in the total entropy of the joint system and heat reservoir. Then the corresponding inequality is given by  
\begin{equation}
R_1=\left\langle \Delta S(x)+\frac{Q[x(\tau);\lambda_y(t)]}{T}- \Delta I(x:y)\right\rangle\ge 0,
\label{PM_ineq}
\end{equation}
where $\Delta S(x)=S(x_\tau,\tau)-S(x_0,0)$ and $\Delta I(x:y)=I(x_\tau:y,\tau)-I(x_0:y,0)$ are changes in Shannon entropy and mutual information. 

When the system is prepared initially in equilibrium and $\lambda_y(t)$ is used for the time-dependent protocol in potential $V(x,\lambda_y(t))$. In this case, one can choose $\widehat{P}(x_\tau |y,\tau)=P_{eq}(x_\tau;\lambda_y(\tau))$ that is the equilibrium distribution of the given dynamics due to final protocol $\lambda_y(\tau)$. Then, one can have $R=(W-\Delta F)/T+I_0$ where $W$ is the non-equilibrium work production given as $W=\int_0^\tau dt (\partial V/\partial\lambda_y)\dot{\lambda}_y$, $\Delta F$ the difference in free energy $F(\lambda_y(\tau))-F_0$, and $I_0$ initial mutual information. The FT for this $R$ leads to generalized Jarzynski equality~\cite{horowitz,ponmurugan}, yielding the resultant inequality 
\begin{equation}
R_2=\left\langle\frac{W[x(\tau);\lambda_y(t)]-\Delta F}{T}+I(x_0:y,0)\right\rangle\ge 0~.
\label{work_ineq}
\end{equation}
In this case, an important issue was raised on what is the optimal protocol change $\lambda_y(t)$ for maximum extraction of work and investigated for various systems~\cite{abreu, horowitz_parrando, parrando_natphys}.

Equation~(\ref{work_ineq}) is an alternative expression for the second law of thermodynamics to Eq.~(\ref{PM_ineq}). The two expressions are equivalent when thermodynamic process is quasi-static. Otherwise, the final PDF will not reach $P_{eq}(x_\tau;\lambda_y(\tau))$ at finite time $\tau$ and the two inequalities are different. Indeed, one can show 
\begin{eqnarray}
R_2-R_1&=&\left\langle -\ln P_{eq}(x_\tau;\lambda_y(\tau))+\ln P(x_\tau,\tau)+I(x_\tau:y)\right\rangle\nonumber\\
&=&\left\langle\ln \frac{P(x_\tau |y,\tau)}{P_{eq}(x_\tau;\lambda_y(\tau))}\right\rangle\ge 0 ~,
\end{eqnarray}
where $P(x_\tau |y,\tau)$ is a real conditional PDF for $x=x_\tau$ at $t=\tau$ given $y$ and the bracket denotes the integral over $x_\tau$ and $y$ with the joint PDF $P(x_\tau |y,\tau)P(y)$. $R_2-R_1$ is the average value of the relative entropy for the PDF to the equilibrium PDF, which can be shown to be always positive. In the following, we will examine Eq.~(\ref{PM_ineq}) that covers more general cases with lower bound and makes it more convenient to investigate the role of mutual information change. 

Using feedback control, one may extract work ($\langle W\rangle<0$) and equivalently heat reservoir loses entropy ($\langle Q\rangle/T<0$) in the absence of a cooler reservoir, which leads to the paradox of Maxwell's demon. The paradox can be resolved by mutual information expended ($-\langle\Delta I\rangle>0$) as correlation between system and memory decays through relaxation. However, mutual information may increase in general due to overshooting in feedback. In the following examples, we show  that overshooting is signalled by {\it negative} correlation accompanied by non-monotonous change of mutual information. It occurs when too large parameters are used for the interval between measurements, the error of measurement, and the intensity of feedback protocol, which is not perfectly avoidable in real experiments.

\section{Szilard engine}

\begin{figure}
\centering
\includegraphics*[width=\columnwidth]{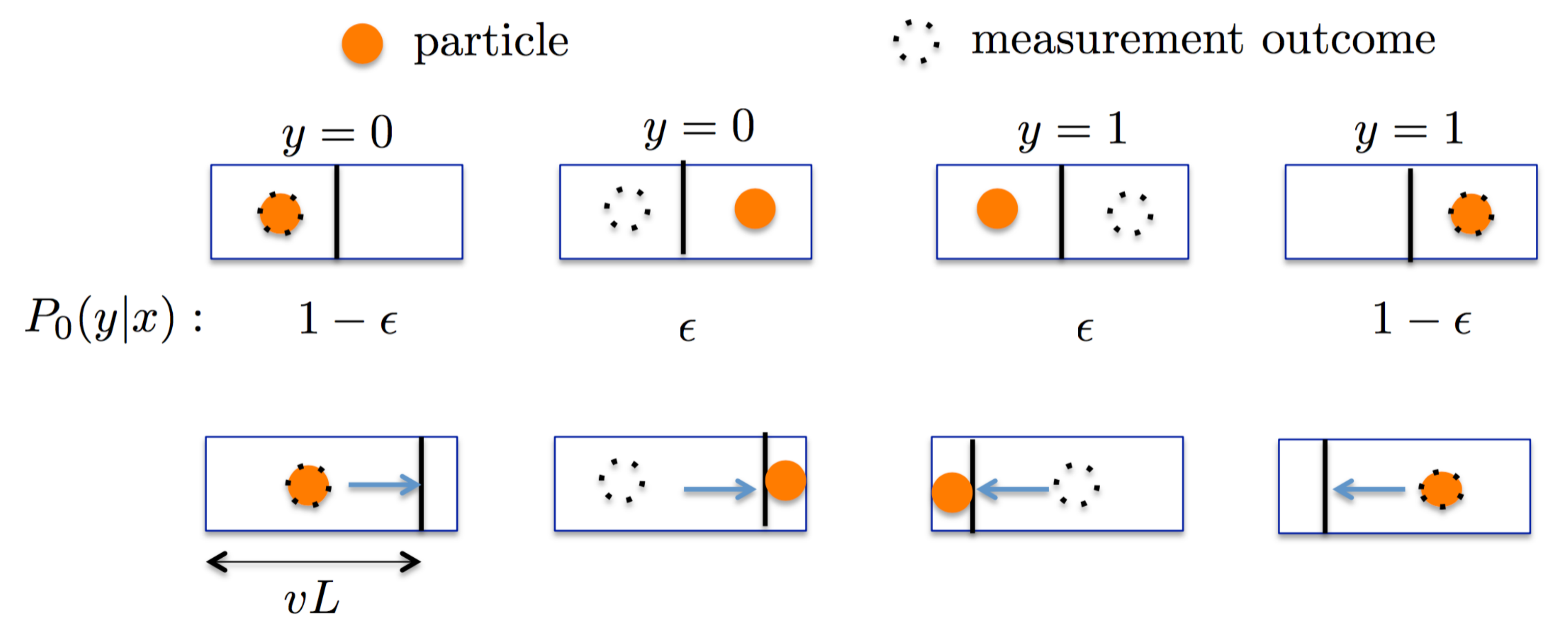}
\caption{(Color online) The Szilard engine. The filled circle denotes the real position of particle  and the open circle the measured position. The first and second rows present four possible situations at the initial moment with measurement probability $P_0(y|x)$ and at an intermediate moment in moving partition.}
\label{fig1}
\end{figure}
We revisit a generalized Szilard engine suggested in a recent study~\cite{sagawa2} focusing more on  mutual information change. Consider one-particle gas in a one-dimensional box with size $L$, which is initially in equilibrium. The partition is inserted at the middle of the box and the position $x$ of the particle is measured. Measurement outcome $y$ reads  $0$ ($1$) if the particle is measured to be in the left (right) of the partition. Inaccurate measurement is described by $P_0(y|x)=1-\epsilon$ for correct measurement, otherwise $P_0(y|x)=\epsilon$ for $0\le\epsilon\le 1$. After measurement, the partition is moved quasi-statically in the direction opposite to the measured position, i.e., to the right for $y=0$ and to the left for $y=1$. In this way the demon tries to expand the gas to extract work. The movement of the partition is stopped when the volume expands up to volume fraction $v$ for $1/2< v\le 1$. See Fig.~\ref{fig1} for a visual help. Finally the partition is removed instantaneously so that the gas suddenly fills the box so as to go back to initial equilibrium state. 

The initial PDF for $x$ is given by $P_0(x)=1/L$. After measurement, the joint PDF for $x$ and $y$ is given from $P_0(y|x)P_0(x)$ as
\begin{equation}
P_0(x,y)=\left\{
\begin{array}{rl}
\bar{\epsilon}/L;&0\le x \le L/2~, y=0\\
\epsilon/L;& L/2\le x \le L~,y=0\\
\epsilon/L;&0\le x \le L/2~, y=1\\
\bar{\epsilon}/L;&L/2\le x \le L~, y=1\\
\end{array}
\right.~,
\end{equation}
where $\bar{\epsilon}=1-\epsilon$ and $\bar{v}=1-v$. 
Then, one can easily find that $p(y)=1/2$, the average mutual information $\langle
I_0(x:y)\rangle=\ln2+\epsilon\ln\epsilon+\bar{\epsilon}\ln\bar{\epsilon}$, and the average entropy $\langle S_0(x)\rangle=\ln L$. 

When the partition is stopped, the joint PDF is found as 
\begin{equation}
P_1(x,y)=\left\{
\begin{array}{rl}
\bar{\epsilon}/(2vL);&0\le x \le vL~, y=0\\
\epsilon/(2\bar{v}L);& vL\le x \le L~,y=0\\
\epsilon/(2\bar{v}L);&0\le x \le (1-v)L~, y=1\\
\bar{\epsilon}/(2vL);&(1-v)L \le x \le L~, y=1\\
\end{array}
\right.
\label{joint_PDF}
\end{equation}
Then, the PDF of the particle is found by summing $P_1(x,y)$ over $y$, given as
\begin{equation}
P_1(x)=\left\{
\begin{array}{rl}
\left(\bar{\epsilon}/v+\epsilon/\bar{v}\right)/(2L);&0\le x \le (1-v)L\\
\bar{\epsilon}/(vL);& (1-v)L\le x \le vL\\
\left(\bar{\epsilon}/v+\epsilon/\bar{v}\right)/(2L);&vL \le x \le L\\
\end{array}
\right.
\label{PDF_final}
\end{equation}
From these results, the average mutual information is found from Eq.~(\ref{MU_def}), given as 
\begin{equation}
\langle I_1(x:y)\rangle=\frac{\bar{v}\bar{\epsilon}}{v}\ln \frac{\bar{\epsilon}}{v}+\epsilon\ln\frac{\epsilon}{\bar{v}}
-\left(\frac{\bar{v}\bar{\epsilon}}{v}+\epsilon\right)\ln\left(\frac{\bar{\epsilon}}{2v}+\frac{\epsilon}{2\bar{v}}\right).
\label{MU_1}
\end{equation} 
The average entropy of the particle is also found from $-\int dx P_1(x)\ln P_1(x)$, given as 
\begin{equation}
\langle S_1(x)\rangle=\ln L-\frac{(v-\bar{v})\bar{\epsilon}}{v}\ln\frac{\bar{\epsilon}}{v}-\left(\frac{\bar{v}\bar{\epsilon}}{v}+\epsilon\right)\ln\left(\frac{\bar{\epsilon}}{2v}+\frac{\epsilon}{2\bar{v}}\right)~.
\label{shannon}
\end{equation}
The work done during this period is found from $-T\int dV/V$, given as
\begin{equation}
\langle W\rangle=-T(\ln2+\bar{\epsilon}\ln v+\epsilon\ln\bar{v})~.
\label{work}
\end{equation} 
Note that $\langle Q\rangle=\langle W\rangle$ since it is an isothermal process. Then one can find the total entropy change in the quasi-static process from Eq.~(\ref{PM_ineq}) as
\begin{equation}
\langle\Delta S_{\textrm{tot}}\rangle_{\textrm{qstatic}}=
\left\langle S_1-S_0-(I_1-I_0)+\frac{Q}{T} \right\rangle=0~,
\label{quasi_static} 
\end{equation}
which holds independent of $\epsilon$ and $v$ where Eqs.~(\ref{MU_1})--(\ref{work}) are used.  The equality manifests that the process is quasi-static, which confirms the legitimacy of Eq.~(\ref{PM_ineq}) as the generalized second law. 

The work extraction is equal to $-\langle W\rangle$. The optimal protocol $v=v^*$ for the maximum work extraction $W^*$ for given $\epsilon$ is determined by extremizing Eq.~(\ref{work}). We get $v^*=1-\epsilon$ and $W^*=T(\ln2+\epsilon\ln\epsilon+\bar{\epsilon}\ln\bar{\epsilon})$. This result was already found in the previous study~\cite{sagawa2}. Here, we discuss this optimal case in more detail. For erratic measurement, expansion should be stopped at an intermediate position for the maximum extraction of work. Interestingly, the maximum work extraction is equal to the initial average mutual information, $W^*=T\langle I_0\rangle$. This is only possible when the initial and final PDFs are the same and the final average mutual information vanishes, as can be seen in Eq.~(\ref{PM_ineq}). It is in fact true, as shown from Eqs.~(\ref{MU_1}) and (\ref{shannon}). For further expansion with $v>v^*$, the amount of work extraction will decrease. The optimal protocols were investigated for other systems such as a two-level system~\cite{abreu}, a Brownian engine operated by time-dependent protocol in potential with feedback ~\cite{horowitz_parrando} and without feedback~\cite{parrando_natphys}.

Mutual information is usually expected to decrease as correlation between system and memory decreases through thermal relaxation. However, it is found to increase after reaching minimum value equal to $0$. We assign values $C$ for correlation between system and memory such as $C=1$ for $y=1~(0)$ and $L/2< x< L$ ($0<x<L/2$), and $C=-1$ for $y=1~(0)$ and $0<x<L/2$ ($L/2<x<L$). Then, its average value defines correlation function $C(x, y)$. From Eq.~(\ref{joint_PDF}), we get 
\begin{equation}
C(x, y)=\frac{\bar{v}\bar{\epsilon}}{v}-\epsilon.
\end{equation} 
$v$ can be regarded as time parameter in quasi-static process. Then, average mutual information $\langle I(x:y)\rangle $ in time is found from Eq.~(\ref{MU_1}) extending $v$ to $1$. Figure 2(a) shows that $\langle I(x:y)\rangle$ decreases for $C(x,y)>0$ while it increases for $C(x,y)<0$. At $v=\bar{\epsilon}$ giving maximum work extraction, $\langle I(x,y)\rangle$ has minimum zero value and $C(x,y)$ is also equal to zero. Negative correlation is possible for nonzero $\epsilon$, i.e., for imperfect measurement which overshoots protocol such that it does not expand, but compress the gas and hence cannot extract work. Note that maximum work extraction $W^*$ is obtained for $v=\bar{\epsilon}$ above which work extraction decreases as compression becomes more probable. 

In the process of free expansion at an intermediate value $v<1$, there are no heat and work produced. Since the engine returns to the initial state, 
there is no mutual information left and no entropy change of the particle. Then, the total entropy change for free expansion is given as
\begin{equation}
\langle\Delta S_{\textrm{tot}}\rangle_{\textrm{free}}= \langle S_0-S_1+I_1\rangle=\left[\bar{\epsilon}\ln\frac{\bar{\epsilon}}{v}+\epsilon\ln\frac{\epsilon}{\bar{v}}\right]>0
\label{free_exp}
\end{equation}
where the inequality holds for all $\epsilon, ~v$. The total entropy increase implies that it is a irreversible nonequilibrium process.

\begin{figure}
\centering
\includegraphics*[width=\columnwidth]{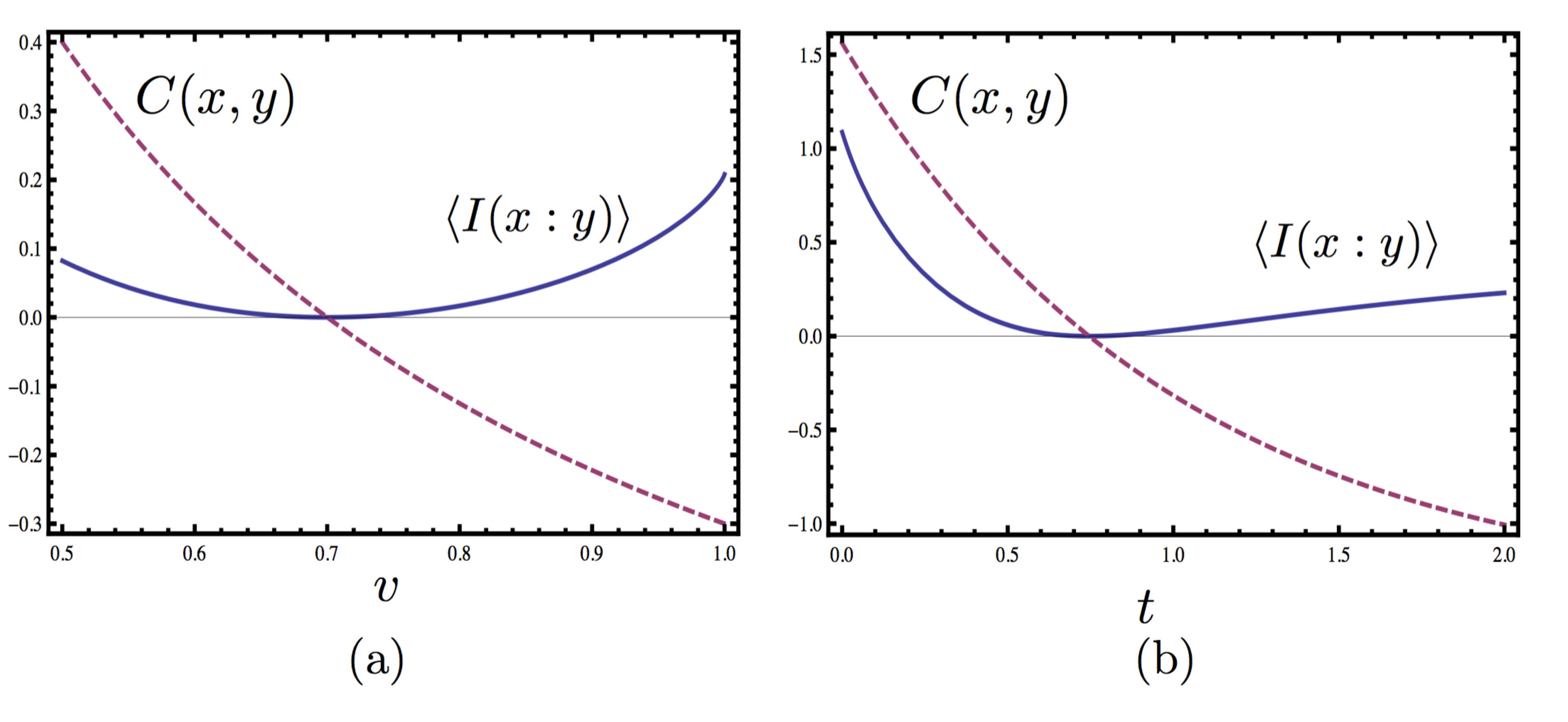}
\caption{(Color online) Mutual information and correlation function: (a) Szilard engine for $\epsilon=0.3$ and $1/2\le v\le 1$. (b) Trap feedback for $\Delta=2$, $k'=0.8$, $\sigma=0.2$, and $0\le t\le 2$. Mutual information decreases to zero as correlation function in the first stage, but increases as negative correlation is built up due to overshooting.} 
\label{fig2}
\end{figure}

\section{trap feedback}

We consider a feedback process where measurement outcome is used for an effective field to restrict the motion of a particle. This so-called cold damping has been studied in various cases~\cite{khkim,jourdan,ito,munakata1}. It was designed to reduce the speed of a particle, while it is very hard to measure fast varying velocity of the particle in experiments usually prepared in overdamped limit. Instead, we investigate the feedback process where the position of the particle is measured and its measurement outcome is used for an effective field to confine the position of the particle toward the center of optical trap. We consider a charged colloidal particle in an optical trap with a harmonic potential. 

We consider multi-feedback process with repeated steps in a time interval $\Delta$. In the beginning of step $i$, a measurement outcome $y_i$ is obtained for an initial position $x_i$ of the particle and electric field $E=-k'y_i/q$ for charge $q$ is applied via feedback, remaining fixed until next step. It is a simple case of $\lambda_y(t)=y$. Then, the corresponding Fokker-Planck equation for the overdamped motion in step $i$ for $t_i\le t\le t_{i+1}$ is given  as
\begin{equation}
\partial_t P(x,t)=\partial_x\left[kx+k'y_i+\beta^{-1}\partial_x\right]P(x,t),
\label{FP_eq}
\end{equation} 
where the friction coefficient is set to unity. Through the repetition of steps with a moderate choice of $\Delta$ and the intensity $k'$, $\langle x^2\rangle$ is expected to get smaller than the value by thermal fluctuation $(\beta k)^{-1}$ in the absence of feedback. 

In step $i$, the probability that the position changes from $x_i$ at time $t_{i}$ to $x_{i+1}$ at time $t_{i+1}$ is written as
\begin{eqnarray}
P(x_{i+1}, t_{i+1};x_i, t_i;y_i)&=&P(x_i,t_i)P(y_i|x_i,t_i)\nonumber\\
&&\times P(x_{i+1},t_{i+1}|x_i,t_i;y_i),
\label{prob_step_i}
\end{eqnarray}
where the measurement probability density is given by $P(y_i|x_i,t_i)=e^{-(y_i-x_i)^2/(2\sigma)}/\sqrt{2\pi\sigma}$. The initial PDF at step $i$ can be written as 
\begin{equation}
P(x_i,t_i)=\frac{1}{\sqrt{2\pi C_i}}e^{-x_i^2/(2C_i)},
\end{equation}
which is Gaussian due to linear force, $-kx$. The conditional probability $P(x_{i+1},t_{i+1}|x_i,t_i;y)$, called the propagator, is the solution of the Fokker-Planck equation in Eq.~(\ref{FP_eq}). We find
\begin{equation}
P(x_{i+1},t_{i+1}|x_i,t_i;y_i)= \sqrt{\frac{a_{\Delta}}{2\pi}}e^{-a_{\Delta}\left(u_{i+1}-e^{-k\Delta}u_{i}\right)^2/2},
\end{equation}
where $u_{i}=x_i+(k'/k)y_i$, $u_{i+1}=x_{i+1}+(k'/k)y_i$, and $a_{\Delta}=(\beta k)(1-e^{-2k\Delta})^{-1}$. Eq.~(\ref{prob_step_i}) is Gaussian with three variables, $x_i,~x_{i+1},~y_i$. Integrating it over $x_{i}$ and $y_i$, one can get the PDF at $t_{i+1}$, written as
\begin{equation}
P(x_{i+1},t_{i+1})=\frac{1}{\sqrt{2\pi C_{i+1}}}e^{-x_{i+1}^2/(2C_{i+1})}~,
\end{equation}
where $C_{i+1}=\langle x_{i+1}^2\rangle$ is related recursively with $C_i=\langle x_i^2\rangle$. 

The recursion relation can be found as
\begin{equation}
C_{i+1}=B+AC_i
\label{recursion}
\end{equation}
where $A=\left[(1+k'/k)e^{-k\Delta}-k'/k\right]^2$ and $B=(1-e^{-2k\Delta})(\beta k)^{-1}+(1-e^{-k\Delta})^2(k'/k)^2\sigma$. Then, we can find
\begin{equation}
C_i=A^{i-1}(\beta k)^{-1}+\frac{B(1-A^{i-1})}{1-A}~.
\label{recursion}
\end{equation}
The recursion relation is stable for $A<1$, which leads to the stability condition:
\begin{equation}
\frac{k'}{k}< \coth\left(\frac{k\Delta}{2}\right).
\label{stability}
\end{equation}
For a stable feedback, $C_i$ approaches to a fixed value $C_\infty=B/(1-A)$, given as
\begin{eqnarray}
\lefteqn{C_{\infty}=}\nonumber\\
&&\frac{(\beta k)^{-1}}{(1+k'/k)}\left[\!\frac{1-
\left[1\!\!-\!\!(k'/k)^2\sigma\beta k\right]\!\!e^{-\frac{k\Delta}{2}}\!\!\sinh\left(\frac{k\Delta}{2}\right)}
{1-(1+k'/k)e^{-\frac{k\Delta}{2}}\sinh\left(\frac{k\Delta}{2}\right)}\!\right].
\end{eqnarray}

\begin{figure}
\centering
\includegraphics*[width=\columnwidth]{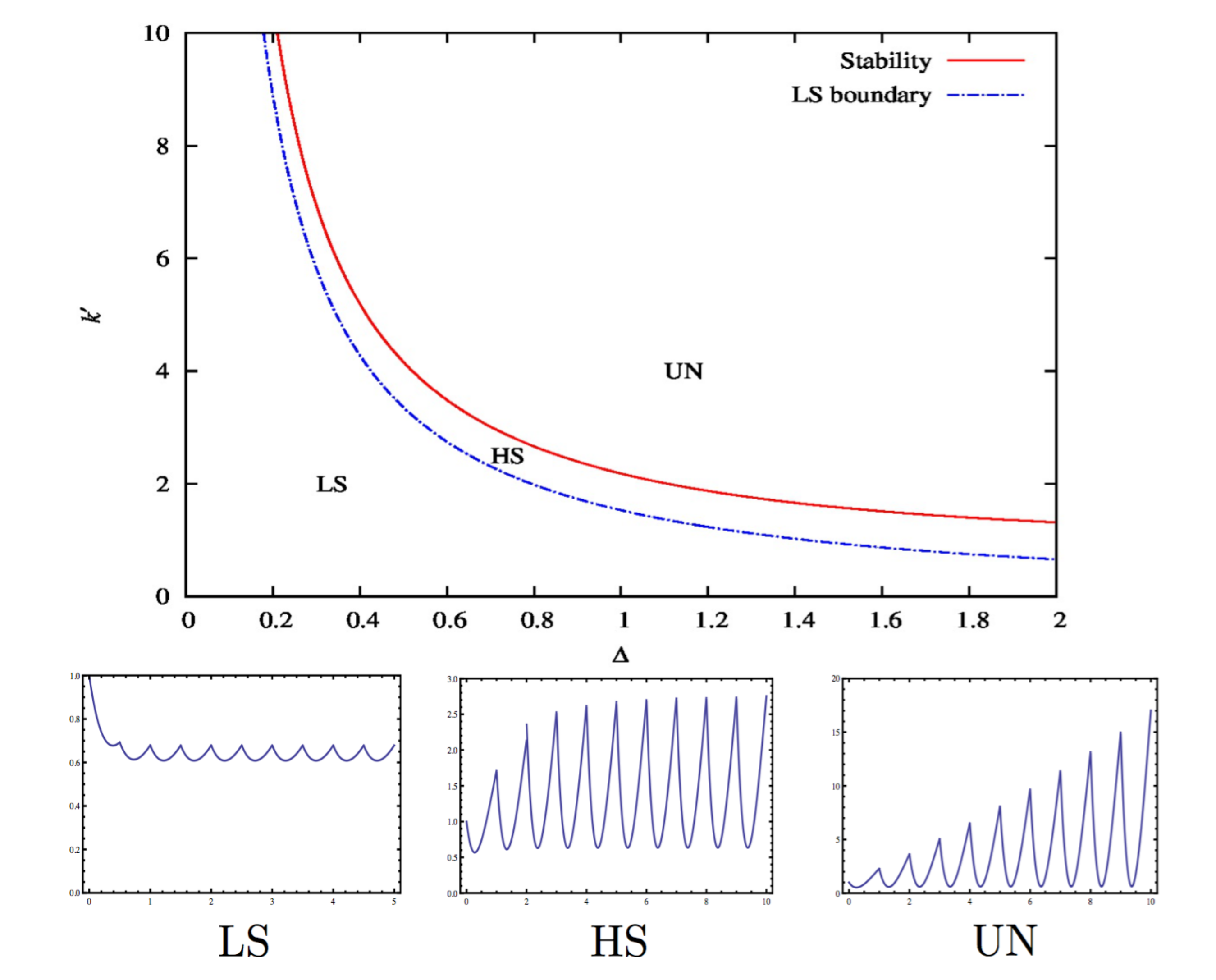}
\caption{(Color online) Phase diagram in $k'$-$\Delta$ plane. Stability boundary for convergent multi-feedback processes is given by $k'/k=\rm{coth}(\gamma\Delta/2)$. LS (HS) denotes the region where  feedback is stable with lower (higher) $C_{\textrm{ave}} $ than $(\beta k)^{-1}$. The boundary of LS region is plotted for $\sigma=0.2$. UN denotes the region with unstable feedback. The three panels below the phase diagram are plots of $\langle x(t)^2\rangle$ (vertical) versus time (horizontal) for LS, HS, and UN regions. The starting values are $(\beta k)^{-1}$, set to unity. The LS curve decays below $1$ and the HS curve increases above $1$ and converges to a finite average value, while the UN curve diverges unlimitedly.}
\label{fig3}
\end{figure}

For large $\Delta $, the effective force $-k'y_i$ may overshoot its restoring role to move the particle toward the center of the harmonic potential. It can be seen from Eq.~(\ref{FP_eq}) that the PDF goes close to the steady state distribution $\propto e^{-\beta k[x+(k'/k)y_i]^2/2}$ due to fast relaxation $\sim e^{-k\Delta}$. As a result, $\langle x^2\rangle=(\beta k)^{-1}+(k'/k)^2 \langle y^2\rangle$ is always larger than thermal fluctualtion $(\beta k)^{-1}$. As steps are repeated, one can find $\langle x^2\rangle =[(\beta k)^{-1}+\sigma (k'/k)^2](1+(k'/k)^2 +(k'/k)^4+\cdots)$, which is even divergent for $k'>k$ away from the stability region. This overshooting behavior also exists in the stability region for finite $\Delta $. Within step $i$ for $t_i\le t\le t_i +\Delta$, $\langle x(t)^2\rangle$ decreases for the first period, but increases for the later period, which can be observed by replacing $\Delta $ by $t-t_i$ in $C_{i+1}$ in Eq. ~(\ref{recursion}) and is shown in Fig.~\ref{fig3}. Feedback is said to be effective if the average value of $\langle x(t)^2\rangle$ for a single interval in $i\to \infty$ limit is lower than $(\beta k)^{-1}$, 
\begin{equation}
C_{\textrm{ave}}=\lim_{i\to \infty}\Delta ^{-1}\int_{t_i}^{t_i+\Delta}dt ~\langle x(t)^2\rangle <(\beta k)^{-1}.
\label{cooling}
\end{equation}

Figure~\ref{fig3} shows the stable region in $k'$-$\Delta$ plane given by Eq.~(\ref{stability}) where the LS region is given by Eq.~(\ref{cooling}) and the HS is for $C_{\textrm{ave}}>(\beta k)^{-1}$. The oscillating behaviors repeated in period $\Delta$ are shown due to overshooting for nonzero $\Delta$ and $\sigma$. In the unstable (UN) region, $\langle x(t)^2\rangle$ is found to increase unlimitedly in time. 

We can compute the total entropy change in Eq.~(\ref{PM_ineq}). 
First, the change in the system entropy during step $i$ is given as
\begin{equation}
\langle\Delta S_{i}\rangle=\left\langle-\ln \frac{P(x_{i+1},t_{i+1})}{P(x_{i},t_{i})}\right\rangle=\frac{1}{2}\ln\frac{C_{i+1}}{C_i},
\label{sys_EP}
\end{equation}
which goes to zero for the stationary stage of feedback with $C_{i+1}=C_i$. 
The entropy production due to the heat production in the heat bath can be found from the thermodynamic first law $-\Delta V+W$ where $V=kx^2/2$, given as
\begin{equation}
\left\langle \frac{Q_i}{T}\right\rangle=\left\langle-\frac{\beta k}{2}(x_{i+1}^2-x_i^2)-\beta k' y_{i}(x_{i+1}-x_i)\right\rangle,
\label{heat_ex1}
\end{equation}
for which the two-point correlation functions from the three-variable Gaussian distribution in Eq.~(\ref{prob_step_i}) are required. Using  
\begin{eqnarray}
\langle y_ix_i\rangle&=&C_i \\
\label{y_var}\langle y_i^2\rangle &=& \sigma+C_i \\
\langle y_ix_{i+1}\rangle&=&C_i\left[e^{-k\Delta}-\frac{k'}{k}(1-e^{-k\Delta})(1+\sigma C_i^{-1})\right],
\label{corr}
\end{eqnarray}
we find
\begin{eqnarray}
\left\langle \frac{Q_i}{T}\right\rangle&=&-\frac{k(C_{i+1}-C_i)}{2T}\nonumber\\
&& +\frac{k'C_i}{T}(1-e^{-k\Delta})\left[1+\frac{k'}{k}(1+\sigma C_i^{-1}) \right]
\label{heat_EP}
\end{eqnarray}

The mutual information can be easily found for a Gaussian PDF given in the form:
\[P(x,y)=\sqrt{\frac{ab-c^2}{(2\pi)^2}}e^{-(ax^2+by^2+2cxy)/2}~.\] 
Note that $a=\langle y^2\rangle/(\langle x^2\rangle\langle y^2\rangle-\langle xy\rangle^2)$, 
$b=\langle x^2\rangle/(\langle x^2\rangle\langle y^2\rangle-\langle xy\rangle^2)$, and 
$c=-\langle xy\rangle/(\langle x^2\rangle\langle y^2\rangle-\langle xy\rangle^2)$. Using 
$P(x)=e^{-x^2/(2\langle x^2\rangle)}/\sqrt{2\pi\langle x^2\rangle}$ and
$P(y)=e^{-y^2/(2\langle y^2\rangle)}/\sqrt{2\pi\langle y^2\rangle}$.  The average mutual information is found as
\begin{equation}
\left \langle\ln \frac{P(x,y)}{P(x)P(y)}\right\rangle =\frac{1}{2}\ln\left[\frac{\langle x^2\rangle\langle y^2\rangle}{\langle x^2\rangle\langle y^2\rangle-\langle xy\rangle^2}\right]~.
\label{MU}
\end{equation}
Then, the change in average mutual information during step $i$ is found as
\begin{eqnarray}
\langle\Delta I_i\rangle&=&\langle I(x_{i+1},y_i)-I(x_i,y_i)\rangle\nonumber\\
&=& \frac{1}{2}\ln\left[
\frac{(\sigma /C_i)}{1+(\sigma/C_i)-(C_i/C_{i+1})H^2}
\right]
\label{mutual_change}
\end{eqnarray} 
where $H= C_i^{-1}\langle y_ix_{i+1}\rangle$, found from Eq.~(\ref{corr}).

\begin{figure}
\centering
\includegraphics*[width=\columnwidth]{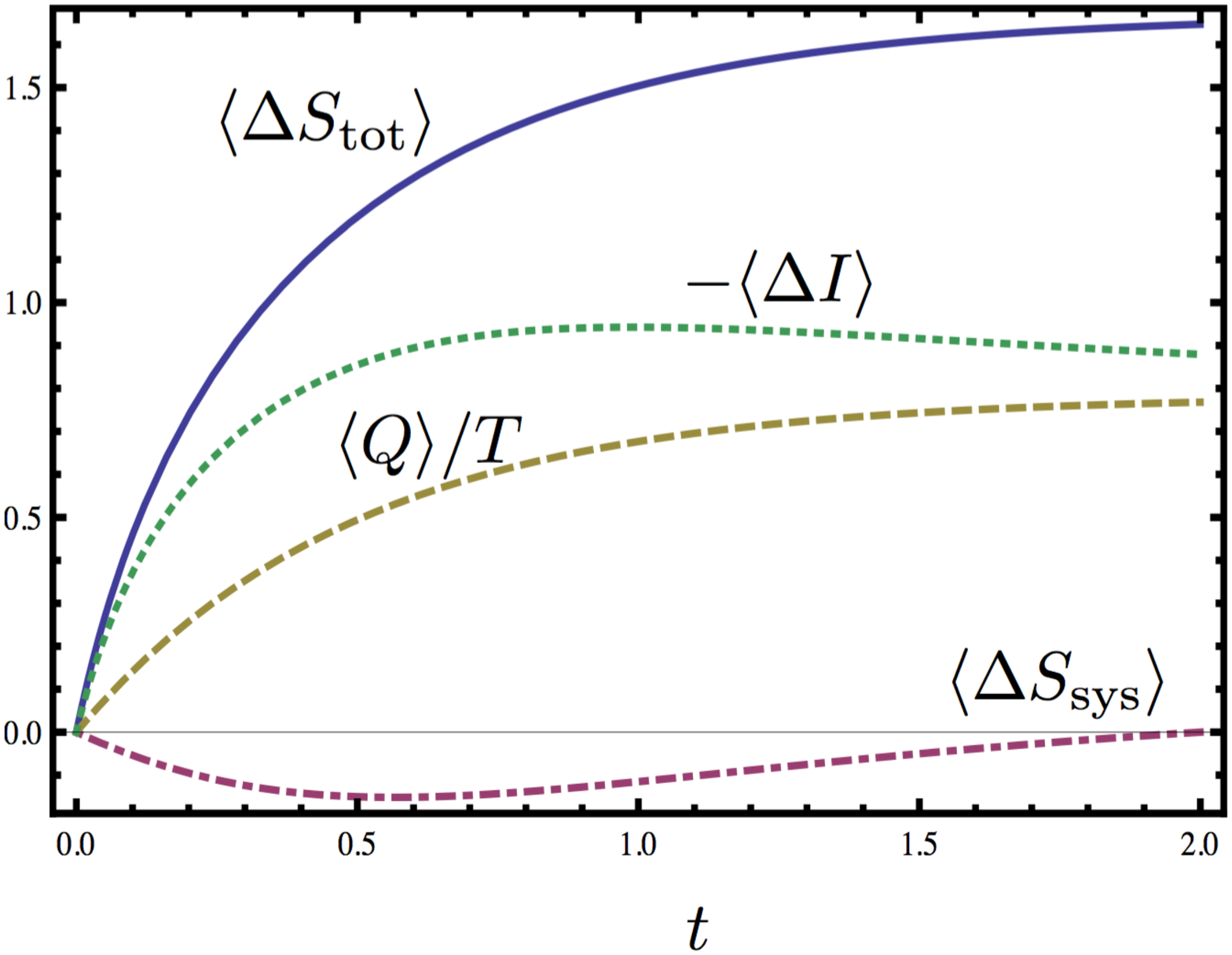}
\caption{(Color online) The change in total entropy $\langle\Delta S_{\textrm{tot}}\rangle$ of the system, memory, and heat reservoir, which is comped of the system entropy change $\langle\Delta S_{\textrm{sys}}\rangle$, heat production in the reservoir divided by temperature $\langle Q\rangle/T$, and minus the mutual information change $-\langle\Delta I\rangle$. The plots are drawn for $0\le t\le \Delta$ in the long-step limit $i\to\infty$ using $\Delta=2$, $k'=0.8$, and $\sigma=0.2$. Total entropy is shown to increase, satisfying the thermodynamic second law.}
\label{fig4}
\end{figure}

Figure 4 shows that the total entropy change always increases. $-\langle \Delta I\rangle$ is positive overall, but has a maximum after which it decreases. As in the Szilard engine, it is due to the fact that mutual information may decrease as negative correlation between system and memory is built up. The correlation function $C(x,y)$ and mutual information $\langle I(x:y)\rangle$ in time can be found from Eq.~(\ref{corr}) and Eq.~(\ref{MU}) by replacing $\Delta $ by $t$. The two functions in the limit  $i\to \infty$ are plotted in Fig.~\ref{fig2}(b) where $\langle I(x:y)\rangle$ has minimum, hence $-\langle\Delta I\rangle$ has maximum, when $C(x,y)$ goes to zero. 

In the limit $\Delta\to 0$, using $t=i\Delta$ in Eq.~(\ref{recursion}), we get 
\begin{equation}
C(t)=\frac{T}{k+k'} \left(1+\frac{k'}{k}e^{-2(k+k')t}\right).
\end{equation}
Then the total entropy production rate is written as $\langle \dot{S}_{\textrm{tot}}\rangle=\langle \dot{S}+\dot{Q}/T-\dot{I}\rangle$, where each rate can be found from Eqs.~(\ref{sys_EP}),  (\ref{heat_EP}), and (\ref{mutual_change}) for $\Delta \to 0$. For steady state with $\dot{C}=0$, we find
\begin{eqnarray}
\left\langle \frac{\dot{Q}}{T}\right\rangle&= &k'\left(1+\frac{\sigma k'}{T}\right),
\label{instant_heat}\\
-\left\langle \dot{I}\right\rangle&=&\frac{T}{\sigma}+k',
\end{eqnarray}
and $\langle \dot{S}\rangle=0$ for steady state. In this continuous measurement limit, $\langle\dot{I}\rangle$ corresponds to information flow, which were previously defined in the framework where both system and memory evolve in time interacting to each other~\cite{horowitz1,horowitz2,shiraishi}.

It is interesting to compare this limit for $\sigma=0$ with the case described by $\dot{x}=-(k+k')x+\xi$ where $-k'y$ is replaced by $-k'x$ for perfect measurement. For $t\to\infty$, the latter goes to an equilibrium steady state with modified temperature $kT/(k+k')$ and there is no entropy production, $\langle \dot{Q}/T\rangle=\langle \dot{W}/T\rangle
\to 0$, while the former goes to a nonequilibrium steady state with nonzero entropy production. For the latter, $\dot{W}$ should be computed by using the Stratonovich calculus~\cite{KYKP} as $-k'(x(t+dt)+x(t))(x(t+dt)-x(t))/(2dt)=-k'(x(t+dt)^2-x(t)^2)/(2dt)$, going to zero in steady state. For the former, however, measurement outcome $y$ is equal to $x(t)$ in perfect measurement, which corresponds to $\dot{W}=-k'x(t)(x(t+dt)-x(t))/dt$ which corresponds to the Ito calculus. It is well known the result depends on the two types of stochastic calculus and the difference is exactly given in Eq.~(\ref{instant_heat}) for $\sigma=0$. 

One intriguing point is that the mutual information change becomes divergent in the limit $\sigma\to 0$, which is also the case for finite $\Delta$, as seen in Eq.~(\ref{mutual_change}). It is indeed due to divergent  mutual information at initial time for perfect measurement of continuous states. From the point of information, it implies that infinite storage space is required to measure continuous states in indefinitely  accurate manner. The generalized second law modified by mutual information change appropriately describes information thermodynamic process by feedback. However, it is not practical to estimate a bound for irreversible quantities such as entropy production and work. For stationary stage of feedback,  $\langle W\rangle > T\langle \Delta I\rangle\to-\infty$ in the limit $\sigma\to 0$.

\section{Summary}
In summary, the generalized total entropy change is found to have a contribution from mutual information change and is shown to be always positive for post-measurement feedback processes. It is shown to be zero for quasi-static expansion of gas of the Szilard engine at any instant before the completion of cycle. For trap feedback, stable and efficient trapping in multi-step feedback process are shown to be possible within a certain range of step-interval and intensity of feedback protocol, which will be a useful information in real experiments. In the limit of perfect ($\sigma=0$) and continuous ($\Delta =0$) measurement, feedback process goes to nonequilibrium steady state, contrary to our expectation based on the effective Langevin dynamics obtained by replacing $y$ with $x$. While divergent mutual information for $\sigma\to 0$ implies well the impossibility of infinite memory capacity for continuous state, it is not useful to estimate the bound for work or entropy production. It will be interesting to find alternative inequalities to Eqs.~(\ref{PM_ineq}) not divergent in $\sigma\to 0$ limit. Time delay in feedback is also an important factor, which cannot be perfectly avoidable in real experiments. We will present the study on the multi-step feedback with time delay in a near future~\cite{KUP}.  

\begin{acknowledgments} 
We thank Professor Takahiro Sagawa for his stimulating suggestions.   
This work was supported by Research fund of Myongji University in 2015.
\end{acknowledgments}

\end{document}